\documentstyle[epsf,psfig]{article}

\newcommand{\up}{\uparrow}
\newcommand{\dw}{\downarrow}

\newcommand{\beq}{\begin{equation}}
\newcommand{\eeq}{\end{equation}}

\begin{document}
\vspace*{0.88truein}

\centerline{\bf A NON-PERTURBATIVE TREATMENT OF THE GENERALIZED}
\vspace*{0.035truein}
\centerline{\bf  SU-SCHRIEFFER-HEEGER HAMILTONIAN ON A DIMER.}

\vspace*{0.37truein}
\centerline{M. COCOCCIONI}
\vspace*{0.015truein}
\centerline{\footnotesize\it Dipartimento di Fisica, Universit\`a di Pavia}
\baselineskip=10pt
\centerline{\footnotesize\it Pavia, I-27100, Italy\footnote{Present address:
International School for Advanced Studies, Via Beirut 2/4, Grignano (TS) I-34014, Italy}}
\vspace*{10pt}
\centerline{\normalsize and}
\vspace*{10pt}
\centerline{M. ACQUARONE}
\vspace*{0.015truein}
\centerline{\footnotesize\it C.N.R.-G.N.S.M., Unita' I.N.F.M., Dipartimento di Fisica,Universita' di Parma}
\baselineskip=10pt
\centerline{\footnotesize\it Parma, I-43100, Italy}
\vspace*{0.225truein}

\vspace*{0.21truein}
\centerline{Abstract.}
Starting from the Hamiltonian for a dimer which includes all the electronic and 
electron-phonon terms consistent with a non-degenerate orbital, by a sequence of displacement 
and squeezing transformation we obtain an effective polaronic Hamiltonian. The renormalized 
electronic interactions differ from the results of semiclassical or perturbative treatments. 
The properties of the variationally determined ground state of two particles in the orbital 
are discussed for variable dimer length in the adiabatic limit.

\section{Introduction.}	
\vspace*{-0.5pt}
\noindent
Thirty years ago, the electron-phonon interaction  
$\gamma^{~}_{ij}\equiv dt_{ij}/d({\bf R}_j-{\bf R}_i)$, 
resulting from the modulation of the hopping amplitude $t_{ij}$ due to the 
displacement $u_i$ of the $i$-th site from its equilibrium position 
${\bf R}_{i}$ in the lattice,
 was indicated as responsible of the BCS superconductivity in transition
metals\cite{labbe}. Later on its relevance also to the physics of 
 quasi-unidimensional conducting polymers\cite{barma} 
and layer compounds\cite{pietro} was recognized. 
Interest in its effects, however, was rather 
sporadic until the seminal paper on dimerisation in unidimensional 
systems\cite{ssh}. Since then, the Hamiltonian 

\begin{equation} 
H_{SSH}=\sum_{\langle i,j\rangle\sigma}
\gamma^{~} _{ij}(c_{i\sigma }^{\dagger }c_{j\sigma}^{{}}
+c_{j\sigma }^{\dagger }c_{i\sigma}^{{}})(u_{j}-u_{i})
\label{Hssh}
\end{equation}
 is known as the Su-Schrieffer-Heeger (SSH) model\cite{fradkin}. The suggestion
was also advanced that it had to do with the bipolaronic 
superconductivity\cite{chackra}, a very exotic
possibility at that time. Its effect in transition metals was reexamined
in Ref.\cite{kuzemsky}. All those issues 
became hot subjects of debate after the
discovery of the high-temperature superconductors, promoting many 
investigations\cite{vari} on its possible effects in the cuprates. 
However, due to the impossibility of an exact
solution on an infinite lattice, the study  of
$H_{SSH}$ has generally been limited to the quasi-classical
approximation, treating $\left( u_{j}-u_{i}\right) $ as a $c$-number, \
and resulting in a modification of the hopping amplitude $t_{ij}$ linearly
depending on  $\gamma^{~} _{ij}$: $t_{ij}^{\ast
}=t_{ij}-\gamma _{ij}\left( u_{j}-u_{i}\right) $. \ A second- order
perturbative treatment has been recently published\cite{zheng1} 
in the context of the equivalence between spinless
SSH interaction term and the {\it X-Y\ }spin- Peierls Hamiltonian,
suggesting a quadratic dependence of \ $t_{ij}^{\ast }$ on $\gamma _{ij}$. \
 On
small size clusters the problem allows in principle 
for an explicit analytical solution with $\left( u_{j}-u_{i}\right) $  properly quantized. 
 Here we shall
present a non-perturbative study of a two-site model system (a dimer) where
the assumed set of electron-phonon interactions generalizes $H_{SSH}$ 
by including other terms
consistent with one electron in a non-degenerate orbital, as discussed 
in Ref.10.\par

\section{The model.}
\noindent
Let us consider  a general two-site electron-phonon Hamiltonian
$H=H_{el}+H_{ph}+H_{el-ph} $ where, in standard notation for a non-degenerate orbital\cite{acquarone}:
\[
H_{el}=\epsilon \sum_{\sigma }(n_{1\sigma }+n_{2\sigma})+\sum_{\sigma}
[t+X(n_{1-\sigma }+n_{2-\sigma})](c_{1\sigma }^{\dagger }c_{2\sigma }+H.c.)
+U(n_{1\uparrow}n_{1\downarrow}+n_{2\uparrow}n_{2\downarrow})
\]
\begin{equation}
+(V-J/2)n_1n_2-2J\left[S^z_1S^z_2+{{1}\over{2}}(S^+_1S^-_2+H.c.)\right]
+P(c^{\dagger}_{1\uparrow}c^{\dagger}_{1\downarrow}
c^{}_{2\downarrow}c^{}_{2\uparrow}+H.c.).
 \label{helbare}
\end{equation}
 and 
$H_{el-ph}$ is the adiabatic Hamiltonian  introduced in Ref.10, which  
 generalizes the SSH Hamiltonian.
\beq
H_{ep}=
g_{\epsilon}^{(12)}\sum_{\sigma}(n_{1\sigma}+n_{2\sigma})(u_{2}-u_{1})
 +\gamma^{(12)}\sum_\sigma (c_{1\sigma}^\dagger
c^{~}_{2\sigma}+c_{2\sigma}^\dagger c^{~}_{1\sigma})(u_2-u_1).
\label{Hadiab}
\eeq
For symmetry reasons, the deformation $u_{i}$ on site $i$ 
at ${\bf R}_i=[(-1^i)a/2,0,0]$ $(i=1,2)$ has to be along the $x$-axis,
therefore $u_{2}-u_{1}$ is the variation of the dimer length $a$, while 
$u_1+u_2=0$. By assuming a
single phonon frequency \ $\Omega$, quantization is performed as usual 
by writing  $u_{i}=L\left(b_{i}^{\dagger }+b_{i}^{{}}\right) $  
with $L=\sqrt{\hbar /2M\Omega }$. It is convenient to 
introduce the odd-parity phonon operators $d,d^{\dagger }$ 
according to $b_{1}^{(\dagger)}-b_{2}^{(\dagger) }=\sqrt{2}d^{(\dagger) }$.
 The even-parity operators do not couple to the electrons, and contribute to the energy a constant term.
 Then, after rescaling the
coupling constants by $\hbar \Omega/\sqrt{2}L $, $H_{ep}$ reads:
\beq
H_{ep}=\left[
g_{\epsilon}^{(12)}\sum_{\sigma}(n_{1\sigma}+n_{2\sigma})
 +\gamma^{(12)}\sum_\sigma (c_{1\sigma}^\dagger
c^{~}_{2\sigma}+c_{2\sigma}^\dagger c^{~}_{1\sigma})\right](d^\dagger+d).
\label{Hadiab1}
\eeq
while free oscillator term reads   $H_{ph}=\hbar \Omega\left(d^\dagger d+1/2\right)$
To obtain an effective electron-only Hamiltonian, describing
fermions dressed by phonons, we perform on $H_{el}+H_{ep}+H_{ph}$ a unitary 
''displacement '' transformation $\exp \left( \delta R\right) $, where
 (from now on we drop the site indexes on $g^{(12)}$ and $\gamma_{12}$ for short):
\begin{equation}
R=\left[ g\left( n_{1  }+n_{2  }\right)
+\gamma\left( c_{1  }^{\dagger }c_{2  }^{{}}
+c_{2}^{\dagger }c_{1}^{{}}\right) 
\right] \left( d^{\dagger }-d\right) =-R^\dagger
\end{equation}
with $\delta $  an up to now undetermined  parameter.
 The "displaced" Bose operators 
$D^{(\dagger)}_i\equiv 
e^{\delta R}d^{\left( \dagger \right) }e^{-\delta R}$ are:

\begin{equation}
D^{(\dagger)}_i
=d_i^{\left( \dagger \right) }-\delta \left[ g\left( n_{1 
}-n_{2  }\right) +\gamma\left( c_{1  }^{\dagger }c_{2 
}^{{}}+c_{2  }^{\dagger }c_{1  }^{{}}\right) \right] 
\label{dim19}
\end{equation}
For the "displaced" Fermi operators $f_{i  }^{\dagger }\equiv e^{\delta R_{\delta
}}c_{i  }^{\dagger }e^{-\delta R_{\delta }}$ we have that the equations of motion (EOM): 
\begin{equation}
\begin{array}{l}
\partial f_{1  }^{\dagger }/\partial \delta=\left(
gf^\dagger_{1\sigma}+\gamma f^\dagger_{2\sigma}\right) (d^{\dagger }-d) \\ 
\partial f_{2  }^{\dagger }/\partial \delta =\left(
gf^\dagger_{2\sigma}+\gamma f^\dagger_{1\sigma}\right) (d^{\dagger }-d)
\end{array}
\qquad   \label{dim23}
\end{equation}
can be decoupled due to
the finiteness of the system, and an analytical solution is obtained. 
Taking into account the 
boundary conditions for  $\delta =0,$  defining for short 
 $B=\delta (d^{\dagger }-d)=-B^\dagger$ 
 and indicating by 
${\rm Ch}\left( x\right) $ and ${\rm Sh}\left( x\right) $ the hyperbolic
cosine and sine, the solutions can be written: 
\[
f_{1  }^{\dagger }=\exp(g B)\left[c_{1  }^{\dagger }{\rm Ch}\left(
\gamma B\right)  +c_{2 }^{\dagger } {\rm Sh}\left( \gamma B\right)\right]
\]

\begin{equation}
f_{2  }^{\dagger }=\exp(g B)\left[c_{1  }^{\dagger }{\rm Sh}\left(
\gamma B\right)  +c_{2 }^{\dagger } {\rm Ch}\left( \gamma B\right)\right]
\label{fermitras}
\end{equation}
One also has $f_{i  }^{{}}=\left( f_{i  }^{\dagger }\right) ^{\dagger 
}$ $(i=1,2)$. Applying 
  Eqs.\ref{dim19} and \ref{fermitras} to $H_{el}+H_{ep}+H_{ph}$ 
yields the ''displaced'' Hamiltonian, 
still containing both Fermi and Bose operators. At this point
 we make the approximation of
factorizing the true wave function $|\Psi \rangle $  into the product of
Fermi and Bose functions, i.e. $|\Psi \rangle =$\ \ $|\Psi_B \rangle
|\Phi_F \rangle.$ The Bose operators are now eliminated from
the ''displaced'' Hamiltonian by averaging $e^{\delta R}He^{-\delta R}$ over a squeezed
phonon wavefunction $|\Psi_B\rangle\equiv\exp(-S)|0_{ph}\rangle $ with
$S\equiv \alpha (d^{\dagger}d^{\dagger }-d^{{}}d^{{}})$ and $d^{{}}|0_{ph}\rangle =0.$ 
In the resulting one-particle polaronic Hamiltonian 
$H^\ast\equiv \langle 0_{ph}|e^Se^{\delta R}He^{-\delta R}e^{-S}|0_{ph}\rangle$
 the renormalized interactions read:
\begin{equation}
\begin{array}{l}
\varepsilon^\ast=\varepsilon-\hbar \Omega\delta(2-\delta)(g^2+\gamma^2) \\ 
t^\ast=t-2\hbar \Omega\delta(2-\delta)g\gamma \\
U^\ast=U (3+\tau^4)/4+(4V-J)(1-\tau^4)/16-2\hbar \Omega\delta(2-\delta)g^2 \\
V^\ast=V(7+\tau^4)/8+(4U+3J)(1-\tau^4)/32-2\hbar \Omega\delta(2-\delta)(g^2-\gamma^2) \\
J_z^\ast=J(5+3\tau^4)/8-(U-V)(1-\tau^4)/2+4\hbar \Omega\delta(2-\delta)\gamma^2 \\
J_{xy}^\ast=J(7+\tau^4)/8-(U-V)(1-\tau^4)/2-2\hbar \Omega\delta(2-\delta)\gamma^2 \\
P^\ast=P(9-\tau^4)/8-(U-V)(1-\tau^4)/4-2\hbar \Omega\delta(2-\delta)\gamma^2 \\
X^\ast=\tau X-2\hbar \Omega\delta(2-\delta)g\gamma 
\end{array}
\qquad   \label{rencf}
\end{equation}
where we have defined 
$\tau\equiv \langle 0_{ph}|e^S{\rm Ch}(GB)|e^{-S}|0_{ph}\rangle
=\exp [-2\gamma^{2}\delta ^{2}\exp (-4\alpha )]$, and we distinguish the longitudinal ($J^\ast_z)$ 
and transverse ($J^\ast_{xy})$ magnetic couplings. 
Our results differ qualitatively under several aspects 
from those following quasi-classical or
perturbative treatments\cite{barma}$^-$\cite{ssh}$^,$\cite{chackra}$^-$\cite{vari}. 
First, we find that 
not only the hopping amplitude $t$, but all the interactions, are non-linearly renormalized. 
Second, if only $\gamma$ is assumed as
 non-vanishing\cite{ssh} then
$t^{\ast }=t$ i.e. {\it in the dimer the $SSH$ interaction
alone does not modify at all the hopping amplitude.} 
Third, due to  
 $0.\le \delta \le 1.$, the cooperative action of both $g$ and $\gamma$ causes $t^\ast<t(<0.)$. i.e.
 it promotes the charge carriers itineracy.

\begin{figure}
\centerline{\psfig{figure=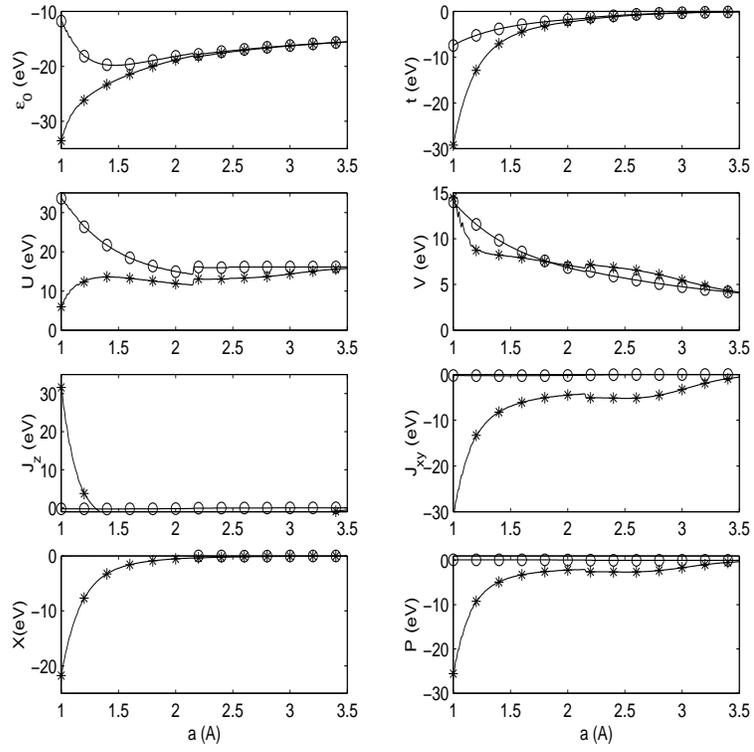,height=10cm,width=10cm}}
\caption {The bare (open dots) and renormalized (asterisks)
 electronic interactions for the case of two particles.} \label{fig1}
\end{figure}
\section{Numerical results.}
\noindent
To study the ground state of the system as the dimer length $a$ is varied, 
we shall adopt  the 
 electron interaction parameters $\epsilon,t,X,U,V,J_z=J_{xy}=P$ as evaluated as in Ref.11, by assuming
 a non-degenerate orbital described
by Wannier functions built from atomic orbitals of Gaussian 
shape. Also the variation with $a$ of $g$ and $\gamma$ has been taken into account, following Ref.10.
 The effective Hamiltonian can be exactly diagonalized\cite{acquarone}yielding the eigenvalues
 as functions of the variational parameters $\delta, \alpha$ and the Wannier functions shape-defining
 parameter\cite{acquarone} $\Gamma$.
  To evaluate the energy of the system, one has to add to $\langle \Phi|H^\ast|\Phi\rangle$, 
the contribution of the squeezed phonons: $E_{ph}=\hbar\Omega[{\rm Sh}(2\alpha)^2+1/2]$. 
From the variational optimization of the total (i.e. electronic plus phononic) energy
 we obtain the true ground state. For lack of space we discuss only the case of two particles
 in the orbital.\par
The bare (open dots) and effective (asterisks) interactions are shown in Fig.1 
for a phonon frequency $\hbar\Omega=0.1$eV. One notices that only 
at low $a$ there is a strong renormalisation in the effective interactions. 
It is due to the fact that, as shown in Fig.2, the  electron-phonon interactions decrease
 with $a$, so that, at low $a$ their renormalizing effect is strong, being weaker for larger $a$.
\begin{figure}
\centerline{\psfig{figure=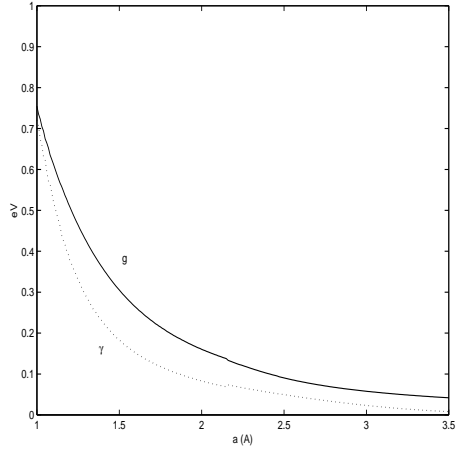,height=6cm,width=6cm}}
\caption {The electron-phonon interactions $g$(full line) and $\gamma$(dotted line) vs. $a$.} \label{fig2}
\end{figure}
The ground state, in the notation of Ref.11, is the singlet state $|Sb\rangle$ for $a <2.16$\AA,
 and the $S=0$ component of the triplet $|T,0\rangle$ for larger $a$. 
At the transition value of $a$, also $\tau$ and $\Gamma$ vary, indicating that the 
 system interactively readjusts both the phononic and the electronic features.

 Knowing the eigenstates $|\Phi\rangle$ of $H^\ast$ we can evaluate the correlation function (CF)
 $\langle \langle Y\rangle\rangle$  
for an operator $Y$ according to 
 $\langle \langle Y\rangle\rangle\equiv
 \langle \Phi|\langle 0_{ph}|e^Se^{\delta R}Ye^{-\delta R}e^{-S}|0_{ph}\rangle|\Phi\rangle$.
 By defining $\theta$ as in Ref.11, one has in the $|Sb\rangle$ state:
\begin{equation}\begin{array}{l}
\langle\langle n_1n_2\rangle\rangle=\cos^2\theta \\
\langle\langle (n_{1\uparrow}^{~}n_{1\downarrow}^{~}+n_{2\uparrow}^{~}n_{2\downarrow}^{~})\rangle\rangle=
\langle\langle (c^\dagger_{1\up}c^\dagger_{1\dw}c^{~}_{2\dw}c^{~}_{2\up}+H.c.)\rangle\rangle=
[1-\tau^4\cos(2\theta)]/2 \\
\langle\langle (S^+_1S^-_2+S^-_1S^+_2)\rangle\rangle=4\langle\langle S^z_1S^z_2\rangle\rangle=
-[1+\tau^4\cos(2\theta)]/2
\end{array}
\eeq
Fig.3 shows the $a$ dependence of the charge distribution CF $\langle\langle n_1n_2\rangle\rangle$,
 of the transverse magnetic CF $\langle\langle (S^+_1S^-_2+H.c.)\rangle\rangle$, from which the
 longitudinal magnetic CF can be deduced by scaling, and of the on-site bipolaron (OSB) CF
 $ \langle\langle (n_{1\uparrow}^{~}n_{1\downarrow}^{~}+n_{2\uparrow}^{~}n_{2\downarrow}^{~})\rangle\rangle$
coinciding with the OSB hopping CF 
$\langle\langle (c^\dagger_{1\up}c^\dagger_{1\dw}c^{~}_{2\dw}c^{~}_{2\up}+H.c.)\rangle\rangle$. 
\begin{figure}
\centerline{\psfig{figure=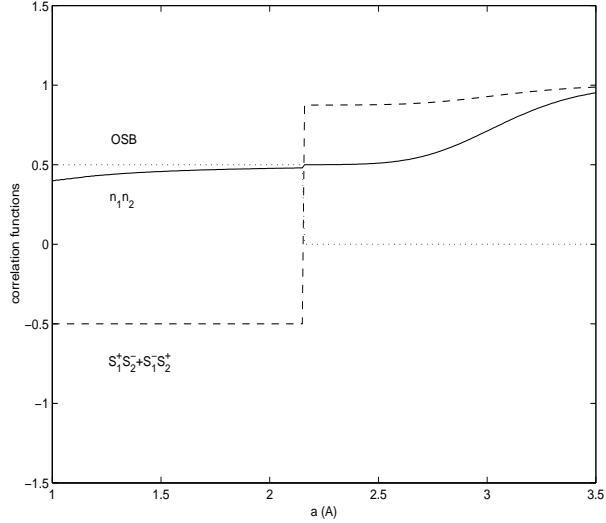,height=7cm,width=8cm}}
\caption {Correlation functions. See the text for explanation.} \label{fig3}
\end{figure}
For $a$ small, where $g$ and $\gamma$ have a strong effect, the formation of an OSB is possible,
 implying an uneven distribution of the charges between the sites, and, at the same time, 
the magnetic CF's are sizeable, even though not fully developed. The phonons therefore induce 
the cohexistence of
 charge and magnetic correlations for dimer lengths such that the electron-phonon
 interactions are not negligible, and promote the itineracy of the OSB.

\section{Acknowledgements}
\noindent It is a pleasure to
thank J.R. Iglesias, M. A. Gusm\~{a}o, M. Cococcioni,
A. Alexandrov, M. Grilli, and particularly A. Painelli,
for critical discussions and comments. This work was supported by
I.N.F.M. and by MURST 1997 co-funded project "Magnetic Polarons in Manganites".\par

\end{document}